# The increase of the ordering temperature of UGa$_2$ under pressure

A.V. Kolomiets[1,2], J.-C. Griveau[3], J. Prchal[1], A.V. Andreev[4], L. Havela[1],

[1] Department of Condensed Matter Physics, Faculty of Mathematics and Physics, Charles University in Prague, Ke Karlovu 5, 121 16 Prague 2, Czech Republic
[2] Department of Physics, Lviv Polytechnic National University, 12 Bandera Str., 79013 Lviv, Ukraine
[3] European Commission, Joint Research Centre, Institute for Transuranium Elements, Postfach 2340, D-76125 Karlsruhe, Germany
[4] Institute of Physics ASCR, Na Slovance 2, 18221 Prague 8, Czech Republic

**Abstract**

Electrical resistivity $\rho(T)$ of the 5$f$ ferromagnet UGa$_2$ was investigated for single-crystal samples as a function of pressure and magnetic field. The Curie temperature monotonously increases from $T_C$ = 124 K under quasi-hydrostatic pressure up to 154 K at $p$ = 14.2 GPa, after which it turns down steeply and reaches $T_C$ = 147 K at $p$ = 15.2 GPa. At 20 GPa the compound is already non-magnetic. This dramatic variation is compatible with exchange interactions mediated by the 5$f$ hybridization with the non-$f$ states. The external pressure first enhances the exchange coupling of the 5$f$ moments, but eventually suppresses the order by washing out the 5$f$ moments. Such a two-band model is adequate for the weakly delocalized 5$f$ states. The spin-disorder resistivity, which produces very high $\rho$-values (300 $\mu\Omega$cm) is gradually suppressed by the pressure. In the paramagnetic state, this leads to a crossover from initial negative to positive d$\rho$/d$T$.

## I. INTRODUCTION

UGa$_2$ crystallizes in the hexagonal AlB$_2$-type crystal structure (space group $P6/mmm$, $a$ = 4.213 Å and $c$ = 4.012 Å) [1] first reported in Ref. 2 and later confirmed by numerous authors. The lattice undergoes orthorhombic distortion in conjunction with magnetic ordering reducing the hexagonal symmetry, [1] which leads to changes not only in the X-ray diffraction pattern but may also affect (as suggested in Ref. 3) the temperature dependence of the electrical resistivity. The application of external pressure of about 16 GPa leads allegedly to a reversible structural transformation to the tetragonal lattice of the Cu$_2$Sb-type with $a$ = 4.648(19) Å and $c$ = 6.316 (36) Å. [4] In the same high-pressure study, the bulk modulus of the AlB$_2$-type was reported to be $B_0$ = 100 ± 7.6 GPa, which is close to the values found for the $R$Ga$_2$ ($R$ = rare-earth) compounds and is more than 1.5 times lower than in ThGa$_2$.

UGa$_2$ is a collinear ferromagnet with Curie temperature $T_C$ = 125 K [1] and magnetic moment of uranium $\mu_U$ = 3.0(2) $\mu_B$ determined by neutron diffraction study, [5] while 2.71 $\mu_B$/U was obtained from the spontaneous magnetization measurements at $T = 4.2$ K.[1] Both $T_C$ and $\mu_U$ are higher than the typical values found for ferromagnetic uranium intermetallics, although $\mu_U$ is still lower than 3.25 $\mu_B$/U or 3.33 $\mu_B$/U, expected for $5f^2$($U^{4+}$) or $5f^3$($U^{3+}$) configurations in the intermediate coupling scheme. The compound shows pronounced magnetocrystalline anisotropy. The [100] (*a*-axis) is the easy magnetization direction, the magnetization along the [120] direction (*b*-axis) is clearly lower, and [001] (*c*-axis) is the hard direction with an estimated anisotropy field of the range 300 T.[1] The powder neutron diffraction has corroborated the in-plane orientation of the uranium magnetic moments, yet without any further specifics about the moment orientation, as well as the non-magnetic state of gallium.[5]

The electronic contribution to the specific heat $\gamma = 10$ mJ/mol·K$^{-2}$ [6,7] is comparable to the value reported for LaGa$_2$, $\gamma = 5$ mJ/mol·K$^{-2}$,[8] and points to the absence of high electronic density of states at the Fermi level. One of the reasons for this could be naturally the 5*f* localization. Polarized neutron data [9] have possibly indicated the $U^{4+}$ state. The localized 5*f* states were suggested by the *ab initio* band structure [10] and CEF calculations,[11] although there remained uncertainty whether they are of the $5f^2$ or $5f^3$ type. The opposite, i.e. itinerant character of the 5*f* states was deduced from photoemission studies.[12] Finally, the magnetoresistance and de Haas-van Alphen experiments [6] could not be convincingly explained by any of the localized or itinerant model if calculating the Fermi surface in the LSDA approximation.

The uncertainty about the character of the 5*f* states was one of the main reasons for the high-pressure studies described in the present paper. The variations of the ordering temperature driven by the lattice compression are an indicator of the nature of the electronic states, which are involved in the magnetic moments formation and exchange interactions. Application of external pressure to a band system typically leads to the suppression of the magnetic moments as well as reduction of the ordering temperatures, resulting from a pressure-induced band broadening. Such negative pressure effect is also connected with the positive magneto-volume relation: the formation of magnetic moments is accompanied by a relatively large increase of the atomic volume and vice versa.[13] Should the magnetic interactions involve localized states, the external pressure would have little effect on the magnetic moments[13] and their ordering temperatures, and in some cases the critical temperature may even weakly increase, due to small changes of the RKKY interaction.

In exceptional cases pressure can support magnetic moment formation. In particular, promoting the smaller magnetic $4f^{13}$ state on the account of larger non-magnetic $4f^{14}$ state in Yb can have such effect. No such effect is, however, expected for U, which is very far from the non-magnetic $5f^6$ state. Nevertheless, a significant increase of the Curie temperature in UGa$_2$ under pressure found in magnetization and electrical resistivity was reported earlier [13-15] but it was based on a relatively limited pressure range of $p \leq 0.8$ GPa. Preliminary data from our present study suggested that the increase extends over much larger pressure range, suggesting the strengthening of the 5*f*-ligand hybridization as the dominant mechanism of the enhancement of exchange interactions.[16] That would classify UGa$_2$ as a material in the interesting regime at the verge of 5*f* localization, which is expected to turn into the standard band ferromagnet only at high pressures. The $T_C$ value has to go in this case through a maximum. On its high pressure side the reduction of $T_C$ is expected to be driven by the U-moments washout.

The possibility to follow the pressure variations variation of the $T_C$ in UGa$_2$ across the whole range of existence of the AlB$_2$ structure type can have therefore the potential to reveal the character of the 5f states in UGa2 and give a guidance for possible theoretical description.

## II. EXPERIMENTAL DETAILS

The measurements were performed on $UGa_2$ single crystals grown by the Czochralski technique. Laue diffraction (Micrometa commercial diffractometer) was used for the quality assessment of the crystals as well as their orientation. A twinning with approximately 2º misalignment of the *a*-axis between the grains was found. Resistivity measurements at ambient pressure were carried out by the four-probe method in the Quantum Design PPMS equipped with a 14 T magnet in the temperature range from 2 - 300 K. The sample size varied from 0.5 mm$^3$ to 2 mm$^3$.

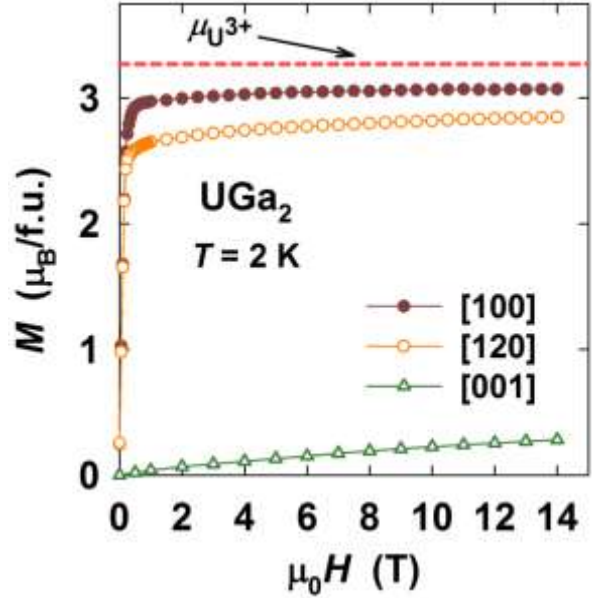

FIG. 1. Field dependence of the magnetization of $UGa_2$ measured at $T = 4.2$ K for different crystallographic directions. The dashed line indicates the magnetization expected for $U^{3+}$.

The high-pressure resistivity measurements were performed by a four-probe DC method in a Bridgman-type clamped pressure cells with a solid pressure-transmitting medium (steatite), using the pressure dependence of the superconducting transition of Pb as a manometer. [18] Before each measurement the cell was loaded and clamped at room temperature. The exact pressure inside the pressure cell was determined later by following the critical temperature of the superconducting transition of Pb. The error bars of the pressure determination are due to the finite width of the superconducting transition of Pb, which is typically 5% to 10 % of the absolute value of $T_C$. In the present experiment that yields an average uncertainty of 0.2 GPa.

Due to a possible change of the sample shape as well as contacts position during pressurization, the absolute values of the resistivity could be calculated only with substantial uncertainty. Therefore, the resistivities measured at high-pressures are presented in arbitrary units. Two different sets of samples were used in these experiments providing independent mutually consistent results.

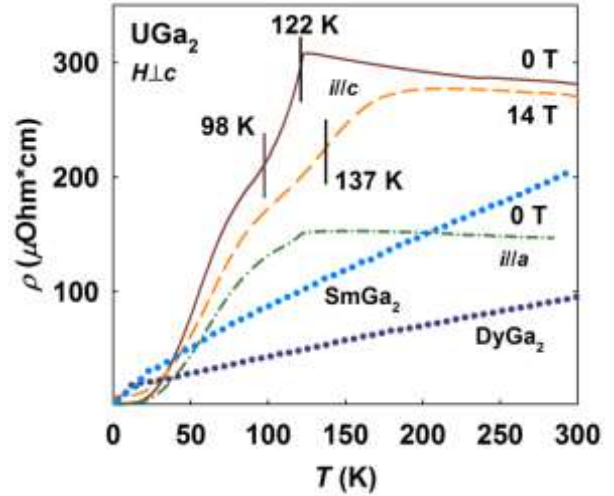

FIG. 2. Resistivity of $UGa_2$ for current *i*//*a* (0 T) and *i*//*c* (0 T and 14 T). Magnetic field $H$ was applied perpendicularly to the *c*-axis. Vertical markers indicate the Curie temperature. The resistivity curves (dotted lines) of $DyGa_2$ and $SmGa_2$ from Ref. 17 are included for comparison.

## III. RESULTS

The quality of as-synthesized single crystals as well as their orientation was verified by magnetization measurements on one of the pieces from the original ingot. The temperature dependence of the magnetization has a shape typical for ferromagnet with a steep drop at the Curie temperature. The $T_C$ value of 124 K determined from the measurement in the field of 0.1 T is in a good agreement with previous literature data. The field dependence of magnetization (Fig. 1) is also typical for a uniaxial ferromagnet with <100> as the easy axis (*a*-axis). The sample used in this study has reached the saturation value of $M_{sat}$ = 3.07 $\mu_B$/f.u. The magnetization for the *b*-axis reaches 2.85 $\mu_B$/f.u. at 14 T but is still far from saturation. The hard-axis magnetization (*c*-axis), is more than 10 times lower in $\mu_0 H$ = 14 T compared to the easy-axis one.

The good quality of single crystal is indicated by the high ratio $R_{300K}/R_{2K}$ = 120. Temperature scans in zero field, $\rho(T)$, performed for the two current orientations *i*//*a* and i//*c*, are shown in Fig. 2. The overall shape and the absolute values of the resistivity are here in a good agreement with the literature data.[3] The striking feature of the UGa$_2$ resistivity is the very high absolute value in the paramagnetic state, exceeding the common Mott limit of metallic resistivity 200 $\mu\Omega$cm especially for the current along $c$[19]. The Mott (or Mott-Ioffe-Regel) limit expresses the fact that the mean free path of conduction electrons cannot be shorter that inter-atomic spacing. In ordinary metals with several conduction electrons per atom and regular spacing of atoms it gives and approximate limit of maximum metallic resistivity, being in the range 100 - 200 $\mu\Omega$cm. It implies that individual contributions to resistivity are not additive any more as the limit is approached, i.e. the Matthiessen's rule of additivity cannot be applied to separate individual contributions. As show by Mooij[20], increasing the resistivity in disordered alloys up to the Mott limit leads to a weak negative slope, d$\rho(T)$d$T$ < 0. Although such effect was originally attributed to special features of density of states, $N(E_F)$, close to the Fermi level, its general occurrence points rather to a weak localization, i.e. quantum interference effect for electron wavelength similar to the spacing of scattering centers, which is gradually disrupted by thermal fluctuations. [21,22]

Such behavior was observed even in actinide alloys, e.g. in simple band systems with low U-U spacing, as U-Mo alloys,[23] which obey the Mott limit restrictions. On the other hand, the class of compounds with larger U-U spacing, which can be vaguely classified as narrow-band systems, exceed the Mott limit considerably. They have often very low residual resistivity (merely 2 $\mu\Omega$cm here), pointing to a weak impurity scattering. As a high resistance appears at high temperatures, sometimes accompanied by a negative slope, [24,25] the reason for high resistivities must be related to the spin-disorder scattering in the paramagnetic state, or to scattering on spin fluctuations in materials, which do not order at all. In such case the negative slope can be also due to the thermal disruption of weak localization, but we do not have any direct evidence.

On the other hand, Kondo effect, which also yields negative d$\rho$/d$T$ at e.g. Ce compounds,[26] should be refused at cases with strongly ferromagnetic ground state and low Sommerfeld coefficient $\gamma$. The Kondo effect yields a strongly correlated non-magnetic or weakly magnetic state with high $\gamma$.

For completeness, one should consider, that the negative d$\rho$/d$T$ can originate from a specific electron-phonon interaction interfering with impurity scattering.[27] The fact that the negative slope in the temperature range above $T_C$ is removed by magnetic field (see Fig.2), which is suppressing the moments fluctuations, means that at least in the case of UGa$_2$ one can consider the decisive effect of spin-disorder scattering and not the electron-phonon scattering.

The resistivity at the maximum, which is located just above $T_C$, reaches 300 µΩcm. Ref. 3 gives even higher value, namely 350 µΩcm, Such disagreement exceeds somewhat the accuracy of determination of the geometrical factor, which should be better than 10%. However, in case of small very small samples, the size of which is restricted by the size of single crystal, which is the case of both the present work and Ref. 3, the difference can be larger.

As the Matthiessen's rule is not valid in the case of UGa$_2$, the resistivity of a non-magnetic analogue cannot be simply subtracted to obtain a pure "magnetic" contribution. The flat $\rho(T)$ in UGa$_2$ in the paramagnetic state means that the electron-phonon scattering has only a relatively small contribution. This is different in rare-earth analogues $R$Ga$_2$ , which are isostructural to UGa$_2$. The $R$Ga$_2$ intermetallics are antiferromagnets with magnetic ordering temperatures not exceeding 15 K and dominating RKKY exchange interaction. $\rho(T)$ curve for DyGa$_2$ is typical for the $R$Ga$_2$ series, and SmGa$_2$ has the highest room-temperature resistivity within the series due to the CEF contribution. [17,29] Neither these two curves, nor any other for the isostructural $R$Ga$_2$ resemble the resistivity temperature dependence of UGa$_2$. The main difference is that lower spin-disorder scattering allows the manifestation of the electron-phonon scattering, responsible for the high temperature linear increase for RGa$_2$.

The contribution to the resistivity due to the spin-disorder scattering in RGa$_2$ is 40 µΩcm or less, despite much higher moments especially for heavy rare earths. [28] This indicates much stronger coupling of the 5$f$ moments to conduction electrons comparing with the 4$f$ counterparts. In general, the ordering temperatures and the spin disorder scattering contribution are 10 times higher in UGa$_2$ compared to $R$Ga$_2$ analogues. On the other hand, in the same compounds the magnetic moments of uranium atoms are about 3 times lower than those of the heave rare-earths. It is known, that for the RKKY interaction both the ordering temperature and spin-disorder scattering scale with $(g-1)^2 J(J+1) J_{sf}^2$ – the product of the squared magnetic moment and $J_{sf}$, the exchange coupling strength between the 4$f$ and conduction electrons. Considering this, one arrives to the conclusion that the effective exchange coupling $J_{sf}$ in UGa$_2$ must be about two orders of magnitude higher than e.g. in DyGa$_2$.

The transition temperature in UGa$_2$ corresponds to the steep drop of electrical resistivity. We have associated $T_C$ with the maximum of d$\rho(T)$/d$T$, which yields $T_C$ = 122 K for $i//c$. The decrease below $T_C$ is slowed down by a hump seen for both current directions. Ref. 6 associates its occurrence with the lattice distortion, but to our opinion there is not real evidence that the distortion would start not at the critical temperature, but 25 K below it. The origin of the hump remains therefore unclear. The Curie temperature for $i//a$ is located at the same temperature as for $i//c$ and corresponds to the similar drop of $\rho(T)$, although it is much less pronounced in the former case and even at ambient pressure it is rather difficult to locate. At low temperatures both resistivity curves drop to very low values, and 1.8 µΩcm and 2.3 µΩcm are achieved at the minimum temperature $T$ = 1.8 K for $i//a$ and $i//c$, respectively.

We can assume that the contribution from the electron-phonon scattering, which follows the $T^3$ dependence, may be small at low temperatures. For instance, Ref. 28 indicates that resistivity of non-magnetic LaGa$_2$ reaches only approx. 10 µΩcm at $T$ = 50 K. We have therefore analyzed tentatively the resistivity curves of UGa$_2$ measured below 30 K in both current orientations using the expression:

$$\rho(T) = \rho_0 + AT^2 + BT(1+2T/\Delta)\exp(-\Delta/T) \qquad (1)$$

The two first terms represent the Fermi liquid approximation of the electron-electron scattering whereas the third term is due to a magnon contribution (electron-magnon scattering). Such type expression should describe $\rho(T)$ well below $T_C$, where magnons as low-energy magnetic excitations dominate, and also where the resistivity values are low with respect to the Mott limit. The parameter $\Delta$ corresponds to the minimum magnon excitation energy (magnon gap), which is the lowest magnetic anisotropy energy – in UGa$_2$ that would be most likely the anisotropy in the basal plane. The rough estimate of the in-plane anisotropy from the magnetization data (Fig. 1) by the linear extrapolation of the [120] $M(H)$ dependence till the crossing point with the [100] curve yields the value of 32 T or, considering the 3 $\mu_B$/U magnetic moment, $\Delta$ of approximately 96 K.

As seen from the Fig. 3(a), the eq. (1) is suitable for the description of the resistivity curves for both current directions $i//a$ and $i//c$. The fitting parameters for the temperature range of 1.8 K < $T$ < 30 K are shown in Table I. They point to a larger contribution of the magnon term for the $i//c$ configuration compared to $i//a$, which is in agreement with generally stronger spin disorder scattering. The spin gap of about 60 K is in a reasonable agreement with the 96 K estimate for the in-plane anisotropy. The Fermi Liquid coefficient $A$ corresponds roughly to the value expected on the basis of Kadowaki-Woods relation $A/\gamma^2 = 10^{-5}$ ($\mu\Omega$cm K$^{-2}$)(mJ/mol K$^2$)$^2$, which yields $A = 1\times10^{-3}$ $\mu\Omega$cm K$^{-2}$ for $\gamma = 10$ mJ/mol K$^2$, but Fig. 3b reveals that the Fermi liquid contribution represents actually only very small fraction of the total resistivity.

Table I. The fitting parameters for eq. (1) applied to the zero-field resistivity of UGa$_2$.

|  | $\rho_0$ ($\mu\Omega$cm) | $A$ ($\mu\Omega$cm×K$^{-2}$) | $B$ ($\mu\Omega$cm×K$^{-1}$) | $\Delta$ (K) |
|---|---|---|---|---|
| | | 0 T | | |
| $i//a$ | 1.7 | 1.1×10$^{-3}$ | 0.9 | 54 |
| $i//c$ | 2.4 | 1.4×10$^{-3}$ | 2.4 | 65 |
| | | 14 T ($H\perp c$) | | |
| $i//c$ | 9.0 | 2.0×10$^{-3}$ | 2.4 | 63 |

The contributions of the electron-electron ($\rho_{FL}$) and the electron-magnon ($\rho_{magnon}$) scattering to the electrical resistivity of UGa$_2$ are illustrated in Fig. 3(b) using the zero-field $i//c$ measurement as the example. It shows that the magnon contribution dominates at temperatures below 35 K, above which the eq. (1) becomes no longer adequate for the description of the experimental data.

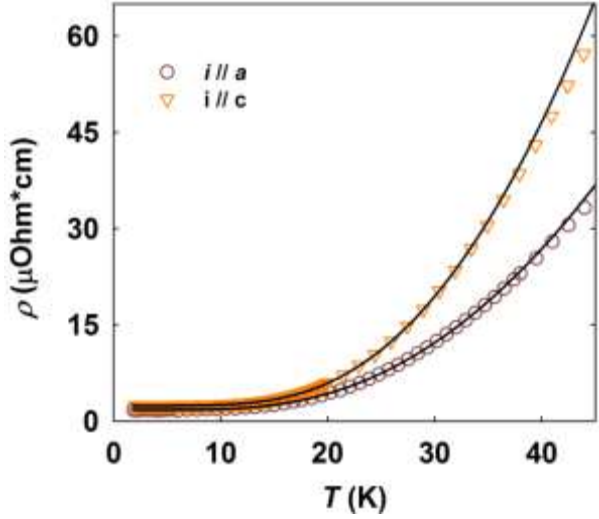 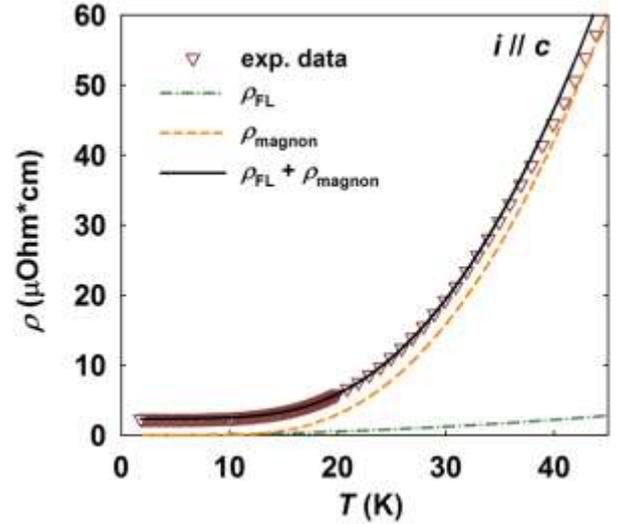

FIG. 3(a). Zero-field low-temperature resistivity of UGa$_2$ for different current configurations. Solid lines represent the fits to the eq. (1). Note: to improve the readability, only 1 point out of 3 is plotted.

FIG. 3(b). Comparison of the Fermi-liquid ($\rho_{FL}$) and magnon ($\rho_{magnon}$) contributions to the zero-field electrical resistivity of UGa$_2$. Note: to improve the readability, only 1 point out of 3 is plotted.

The application of magnetic field perpendicular to the *c*-axis affects the $\rho(T)$ dependence in the whole temperature range (Fig. 2): it increases the resistivity at the low temperature end, whereas the effect at high temperatures is the opposite. The crossing point between these two regions for the field of 14 T is around 36 K. The most pronounced change of the resistivity dependence takes place in the vicinity of the Curie temperature, where the $T_C$ anomaly shifts from 122 K (0 T) to 136 K (14 T). The shift of the anomaly towards the higher temperatures with the increasing field is due to the combination of the two effects: the increase of the Curie temperature and suppression of the of the spin-disorder scattering on the fluctuating U moments. The negative d$\rho$/d$T$ is almost entirely removed, too, and the maximum resistivity drops by the factor of 1/3 or ≈ 100 μΩcm (see Fig. 2). The changes of the residual resistivity can be attributed to the normal magnetoresistance. It increases more than 3 times from 2 μΩcm to 9 μΩcm (Fig. 4). The room-temperature resistivity is affected less by the field increase. It remains nearly constant till 6 T and then continuously decreases by 9% as 14 T is reached. The eq. (1) is still suitable for the description of the 14 T data, and it yields $\Delta$ = 63 K, $\rho_0$ = 9.0 μΩcm, $A$ = 2.0×10$^{-3}$ μΩcm·K$^{-2}$, and the coefficient $B$ remains unchanged within the precision of the fit. The temperature range, across which the eq. (1) can be applied, shrinks from 2 K < $T$ < 30 K in zero field to 2 K < $T$ < 18 K in 14 T.

High-pressure resistivity measurements were performed using only the *i*//*c* current orientation because the anomaly at the Curie temperature is more pronounced compared to the *i*//*a* setup. Hence, the *i*//*c* setup provides more reliable estimate of the $T_C$ for the compressed sample, despite the $\rho(T)$ features are broadened, perhaps due to a non-hydrostaticity.

As seen from Fig. 5(a), the $\rho(T)$ shape gradually evolves with increasing pressure with no sudden jumps or discontinuities. One can clearly recognize that the Curie temperature increases with increasing pressure. This change is well illustrated by the blow-up of the transition region (Fig. 5(b)). The drop of the resistivity associated with the transition is shifted towards higher temperatures with increasing pressure. The analysis of the first derivative for each $\rho(T)$ dependence was used to establish the precise values of $T_C$ at all pressures (Fig. 6). It was found that the Curie temperature of $UGa_2$ increases with increasing pressure till $T_C = 154$ K at $p = 14.2$ GPa. Above this pressure, the

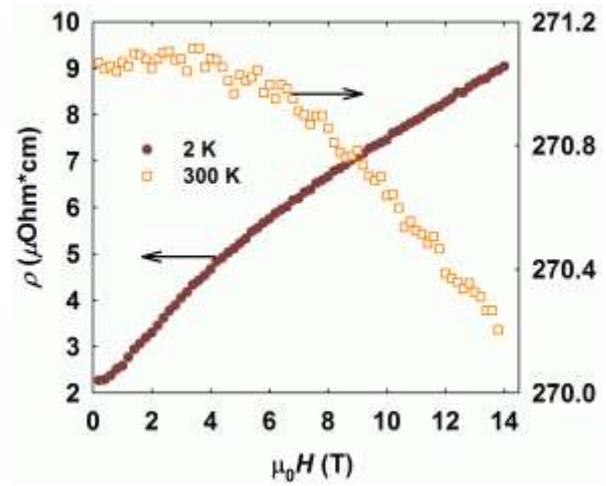

FIG. 4. Field variation of the resistivities measured at 2 K (full circles) and 300 K (open circles) for $UGa_2$ in the $i//c$ configuration. Note two vertical axes.

Curie temperature starts to decrease much steeper than it increased. The closer inspection of the $T_C(p)$ dependence suggests that the maximal value of $T_C$ is reached at about $p = 13$ GPa although there are no experimental data at this pressure value. At $p = 15.2$ GPa, $UGa_2$ is still ferromagnetic with the ordering temperature $T_C = 147$ K.

One of the samples was loaded to the pressure 19.5 - 21 GPa (dotted lines in Figs. 5a and 5b). In such state it did not show any traces of the magnetic phase transition. Enhanced $\rho_0$ could be due to the structural transformation, which was reported around 16 GPa.[6]. Since the contacts to the sample were not lost and its $\rho_{300K}$ value corresponded to the value expected from the $\rho_{300K}(p)$ dependence for the other pressures (Fig. 7), we assume that the suppression of the magnetic order in $UGa_2$ indeed takes place between 15.2 GPa and 21 GPa. Due to the lack of the experimental data in this pressure range the exact value of the critical pressure could not be determined. Still it should be mentioned that the extrapolation of the $T_C(p)$ dependence (Fig.6) to the intercept with the temperature axis suggests that $T_C$ should turn to zero between 16 GPa and 18 GPa. As to the high-pressure structure, Ref. 4 indicated the possibility of continuous modifications of interplanar distances. However, a clear identification of the high pressure structure is necessary in order to establish the correlation between the structure and magnetism of the high pressure phase.

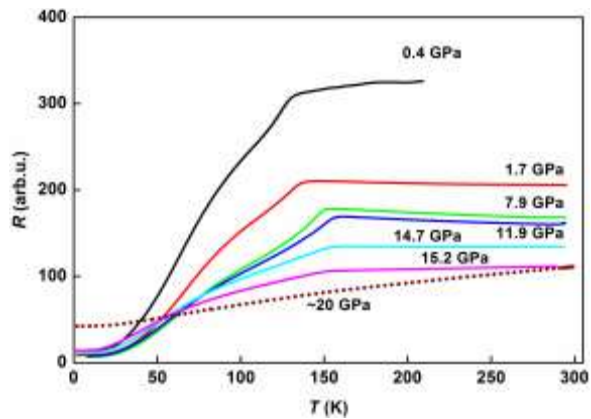

FIG. 5(a). Selected resistivity curves measured of $UGa_2$ measured in the $i//c$ configuration. The dotted line represents the data collected on the sample, which has been exposed to the pressure over 16 GPa.

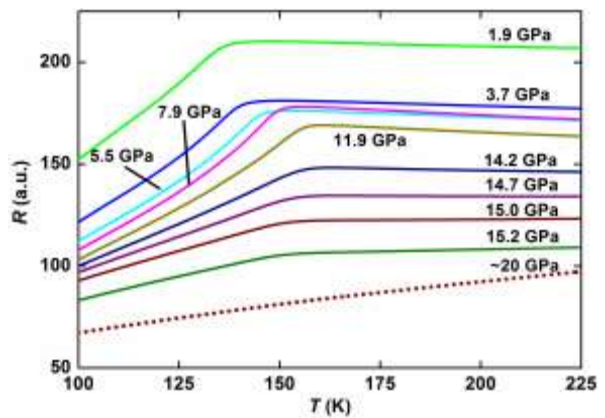

FIG. 5(b). The detail of the resistivity curves of $UGa_2$ in the transition region ($i//c$ configuration). The dotted line represents the data collected on the sample, which has been exposed to the pressure over 16 GPa. (also use arb.u.)

When analyzing $\rho(T)$ obtained in the high-pressure experiment, one should consider the variation of the geometrical factor during the pressure increase, which makes the absolute resistivity values unreliable. Since the contacts are not firmly attached to the sample in this type of cells, we particularly expect small variations in the 1-2 GPa region, whilst at higher pressures the contacts are typically already immobilized. We assume that this is the most plausible reason for the resistivity drop between 0.4 GPa and 1.7 GPa (Fig. 5(a). By comparing the data collected above 1.7 GPa (Fig. 5(a) and 5(b)) we indeed observed a smooth (if any) variation of the geometrical factor in our experiment and, therefore, we consider the pressure-induced changes of the residual and room-temperature resistivity to be intrinsic. Although the contacts can in principle still shift in this pressure range, a typical fingerprint is a large sudden change, which is absent here.

The residual resistivity, which could be reasonably well represented by $\rho(2.5$ K$)$, has the U-shaped pressure dependence with the minimum between 6 GPa and 8 GPa, with a progressive increase on the high-pressure side (Fig. 7). Contrary to $\rho(2.5$ K$)$, the room-temperature resistivity is reduced under pressure (Fig. 7). Since the $\rho(T)$ curves in the pressure cell could be reliably measured only till approximately $T = 250$ K, the value $\rho(250$ K$)$, we use it as an indicator instead of $\rho(300$ K$)$. As seen from Fig. 7, the $\rho(250$ K$)$ decreases in the whole pressure range. It develops through a plateau between 3 GPa and 8 GPa, after which turns downwards and drops steeply above 12 GPa. The similarity of the pressure values, at which both $\rho(2.5$ K$)$ and $\rho(250$ K$)$ change their character, suggests that the phenomena behind their pressure variations may be the same.

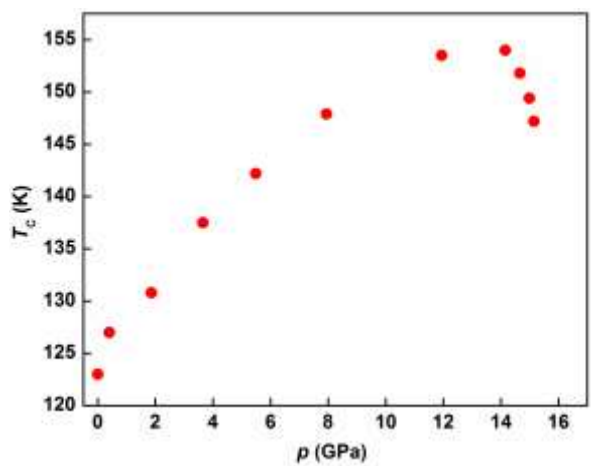

FIG. 6. Pressure dependence of the Curie temperature of $UGa_2$.

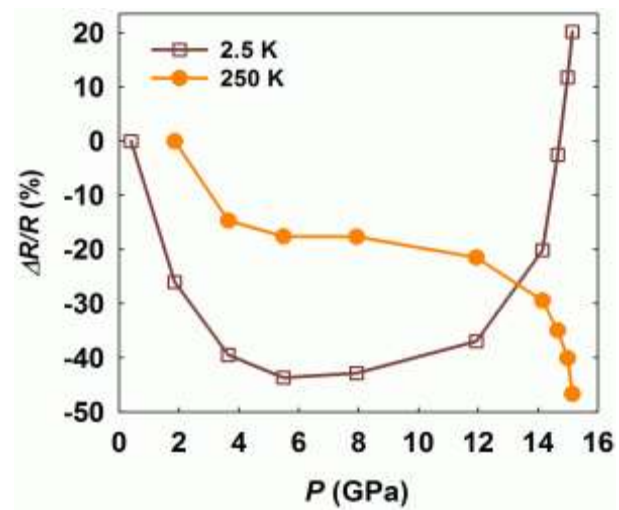

FIG. 7. Pressure effect on the resistivity of $UGa_2$ at $T = 2.5$ K (open squares) and 250 K (full circles). The resistivities were compared with the respective values at $p = 0.4$ GPa. The lines are eye-guides.

The steep variation of both residual and room-temperature resistivities is not the only qualitative change, which takes place around 12 GPa, the character of the $\rho(T)$ dependence changes in the same pressure range. The anomalous decrease of the resistivity with increasing $T$ in the paramagnetic state was found for all pressures below $p = 12.5$ GPa. At $p = 14.7$ GPa a local minimum appears at $T \approx 220$ K. Finally, $d\rho/dT > 0$ at $p = 15.0$ GPa.

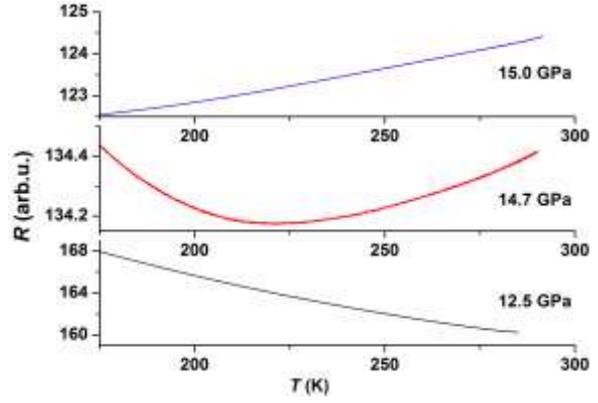

FIG. 8. Positive $d\rho/dT$ in the paramagnetic state developing with pressure.

The low-temperature resistivity of UGa$_2$ can be fitted to the eq. (1) for all pressures. The temperature range, in which the fit is valid, shrinks somewhat with increasing pressure: its upper limit shifts from 35 K at ambient pressure to 20 K at 15 GPa. Still, the agreement between the experimental data and the fit remains quite good. The obtained fitting parameters are shown in Fig. 9. The character of their pressure dependence points to an increase of the magnon band gap. The energy of the magnon band gap is a measure of magnetic anisotropy energy per one U atom, it may be understood as due to strengthening of hybridization, which is an important ingredient of the two-ion anisotropy. Its decrease in the high-pressure range can be related to the suppression of U magnetic moments. The pre-factor $B$ develops very similarly. It contains, besides parameters of the Fermi level, the strength of coupling of the conduction electrons to the spins of 5$f$ states, which is also likely supported by pressure, until the 5$f$ moments plunge. The pre-factor $A$ of the Fermi-liquid term exhibits a weakly increasing tendency, which can testify increasing density of states at $E_F$, also corresponding to a stronger hybridization of conduction-electron states at $E_F$ with the 5$f$ states, which practically do not contribute to $N(E_F)$ at ambient pressure. The uncertainty is, however, high, due to the relatively small contribution of the $e$-$e$ scattering to the total resistivity.

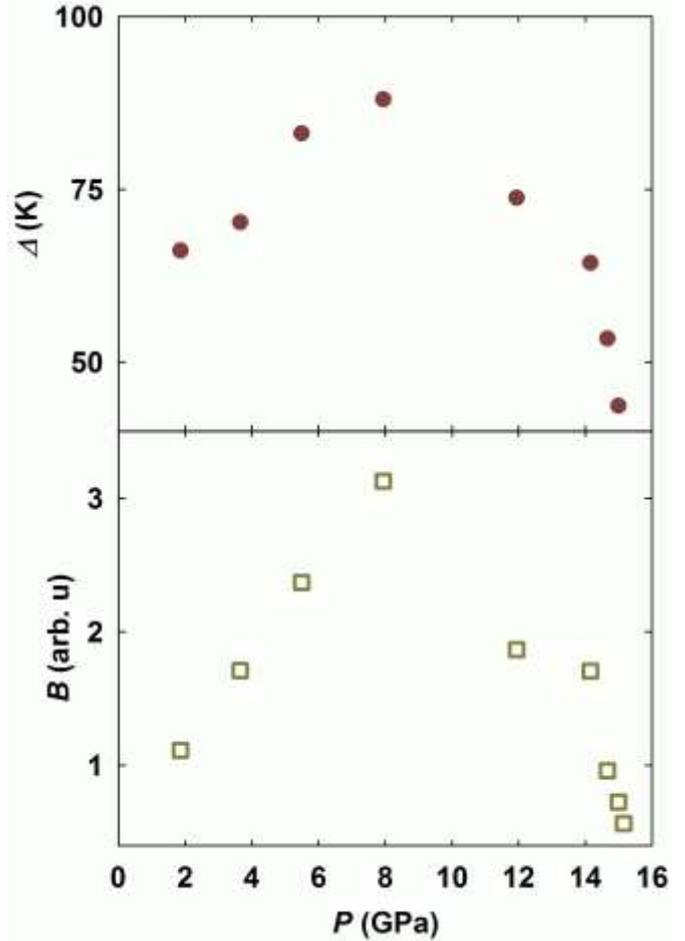

FIG. 9. Pressure variation of the magnon band gap and the prefactor $B$ of the magnon term in eq. (1).

## IV. DISCUSSION

The obtained experimental results indicate that the 5$f$ states of uranium in UGa$_2$ are not common band states. In such case the ordering temperature would decrease with pressure from the beginning. Neither are they fully localized since that would presume very small changes of $T_C$. An intermediate situation is the only scenario consistent with the observed pressure variation of the critical temperature, namely its pronounced increase followed by the saturation and the subsequent drop.

Such non-monotonous behavior can be obtained, for example, in the framework of the Doniach necklace model [30]. In this model, both the RKKY interaction and the Kondo interaction depend on the exchange parameter $J$ between local spins and spins of conduction electrons. Strengthening of the hybridization leads to an increase of $J$, which first increasing the inter-site exchange coupling, but when it becomes too strong the Kondo screening leads to a collapse of local moments and magnetic order vanishes. Such model is often used for e.g. Ce-based materials. It is, however, not quantitavely applicable to strongly ferromagnetic U compounds, at which we cannot expect the applicability of the simple RKKY type of interaction (notice that the ordering temperature of $UGa_2$ is an order of magnitude higher than for $RGa_2$ compounds with RKKY interaction) or the compensation of U moments of 3 $\mu_B$ by a Kondo type screening.
The model, which was so far successfully applied to light actinides, is based on hybridization induced exchange interaction (operating independent on the RKKY mechanism), which is strongly directional. It is also based on mixing two types of states, namely strongly correlated *f*-states and conduction-electron states, conceiving their mixing in terms of a resonant scattering. It extends the Coqblin-Schrieffer theory [31] and provides a strongly anisotropic exchange interaction, with strong ferromagnetic exchange along the shortest links between the *f*-atoms and magnetic moments oriented perpendicular to them, while magnetic anisotropy energies can be very high. [32] In such model, strengthening the hybridization depending on the inter-atomic spacing by pressure at first leads to stronger effective inter-site 5*f*-5*f* exchange interactions. But at a certain stage the hybridization starts to affect the 5*f* moments, which are washed out, and finally the magnetic moments and their ordering is lost in a strongly non-linear way.

The character of the $T_C(p)$ dependence in $UGa_2$ goes in line with the predictions of such two-band approach, in which the hybridization plays a dual role: on one hand it strengthens the 5*f*-5*f* or 5*f*-ligand hybridization, and on the other – it leads to the washout of the 5*f* magnetic moment due to broadening of the 5*f* states. The maximum of $T_C$ can be interpreted just as the point, in which the product of the exchange strength and the squared magnetic moment reach its maximal value. Upon further increase of pressure the strengthening of exchange interactions is overcompensated by the moments washout. It would be interesting to follow the development of magnetization, but pressures around 10 GPa are so far out of the reach of common techniques yielding quantitative magnetization data. So far, such pressure-induced increase of the critical temperature has been reported only for few U compounds, namely for UTe and USe, [33] UPtAl, [34] or $UIn_3$. [35] This makes the case of $UGa_2$ even more interesting. It is, therefore, understandable that standard computational schemes do not give adequate description of the Fermi surface topology. It would be interesting to see if the LDA+U methods, better capturing more localized systems, would explain the experimental observables in dHvA.

The important role of the 5*f*-ligand hybridization in mediating the exchange interaction is evident from the fact that the shortest U-U spacing $d_{U-U}$ = 4.012 Å by far exceeds the Hill limit, preventing a direct 5*f*-5*f* overlap. Even after the 11% volume reduction[4] following the application of quasihydrostatic pressure of 15 GPa the $d_{U-U}$ remains in the range 3.85-3.90 Å, i.e. far above the Hill limit. Here we have assumed the isotropic volume reduction since the individual compressibilities, which could be anisotropic, have not been reported.

It should be pointed out that the situation, in which the 5*f* magnetic moments are forced by anisotropy into the direction perpendicular to the shortest U-U links, is typical for anisotropic hybridization-mediated exchange interaction, [32] although it may have a broader validity in those materials, where the 5*f* states are involved in anisotropic bonding, and strong spin-orbit interaction leads to large orbital moments even in the case of band states. [36] For $UGa_2$ the shorter U-U spacing along the *c*-axis (4.012 Å) compared with the longer one within the basal plane (4.21 Å) implies that the uranium moments are in the plane and a strong ferromagnetic coupling is along *c*. A certain insight is provided by inelastic neutron diffraction data. [37] They indicate the absence of the crystal-field excitations, but show the magnon excitations with dispersion along *c*, confirming that the exchange along *c* is the prominent driving force of the magnetic ordering. The magnon gap of 7-8 meV (80-90 K) is in a reasonable agreement with the gap obtained from $\rho(T)$ and with the moderate *a-b* anisotropy, which allows low energy propagating magnon modes with the U moments perpendicular to *c*. The observation of magnons can serve as additional proof of certain 5*f* localization, as band 5*f* systems do not exhibit clear magnon excitations.[38]

The dominating role of the 5*f*-ligand hybridization can be deduced also from the work of da Silva *et al.* [8], who have shown that the substitution of Si or Ge for Ga has more pronounced effect on the Curie temperature of $UGa_2$ than the volume change of the same magnitude, but without the modification of the electronic character of the ligand.

Finally, the observation of the uranium moments slightly below the $U^{3+}$ or $U^{4+}$ values and the simultaneous absence of any substantial 5*f* density at the Fermi level as indicated by the low electronic contribution to the specific heat $\gamma = 10$ mJ/mol·K$^{-2}$ [6,39] also favors the 5*f*-ligand hybridization over the pure 5*f*-band formation when we consider the main delocalization mechanism of the 5*f* states.

The character of $UGa_2$ in the high-pressure phase existing over 16 GPa remains still unclear, as the structure has not been yet identified unambiguously. Although it was mentioned that the symmetry changes from the hexagonal to the tetragonal [4], while the interplanar distances vary continuously across the transition, there are certain doubts as to the structure details of the high-pressure phase. First, the suggested $Cu_2Sb$-type structure is not among the $AlB_2$-type derivatives.[40] Second, the lattice parameters for the high-pressure $Cu_2Sb$-type phase reported in the original work[4] give higher volume per 1 formula unit of $UGa_2$ than the ambient pressure $AlB_2$-type hexagonal phase. To our opinion, that calls for further investigation of the structural transition and its role in the suppression of magnetism in $UGa_2$.

The reported results indicate a special position of $UGa_2$ among U compounds. A proximity to the localization of 5*f* electronic states is evidenced by the fact that simple 5*f* band picture is not compatible with the increase of $T_C$ under pressure. In such situation it is understandable that the conventional DFT calculations cannot explain the Fermi surface geometry, which affects the dHvA data. [3] It is a matter of fact that majority of U intermetallics exhibit the 5*f* band states at the Fermi level. The only known exception, at least among binary compounds, is $UPd_3$, where the localized 5*f* states displaced from the Fermi level were clearly identified by photoelectron spectroscopy. There are several other compounds with low $\gamma$-value, e.g. UPdSn with $\gamma = 5$ mJ/mol K$^2$. [41] However, both dilution studies [42] and spectroscopy data [43,44] speak against the localization. Calculations in this case succeeded to capture the Density of States if performed for the real non-collinear magnetic structure. [45] A pressure dependence of the ordering temperature is positive in the case of UPdSn, but the increase is slower than in in UGa$_2$ (1.4 K/GPa) and the studied pressure range is limited so the maximum value of $T_N$ is unknown. [46] Also the bulk modulus is unknown and the variation of $T_N$ cannot be therefore related to the absolute lattice compression parameters.

Another interesting case, helping to determine the location of UGa$_2$ on the landscape of U intermetallics, is UIn$_3$, mentioned above. It is a local moment antiferromagnet, $T_N$ = 88 K, which increases to 127 K for $p$ = 9 GPa (i.e. 3 K/GPa). Higher pressures could not be reached. [35] This can be compared with the isostructural UGa$_3$, a 5$f$ band antiferromagnet with $T_N$ = 67 K, the ordering of which is suppressed by the pressure 2.5 GPa. [47] When the size of ligand, which strengthens the 5$f$-ligand hybridization, is reduced even more in UAl$_3$, the magnetic ground state is suppressed even at ambient pressure, and UAl$_3$ is a weakly temperature dependent paramagnet.[48] Unfortunately only basic data are known in the opposite case, UTl$_3$, where the hybridization should be the weakest from the series. U moments are larger (1.6 $\mu_B$) than in UIn$_3$ (1.0 $\mu_B$) (Ref. 49 and references therein) although the ordering temperatures are comparable. Variations of $T_N$ with pressure are not known.

Considering such facts, we may speculate that the strong increase of ordering temperatures up to a maximum and subsequent downturn may represent a generic type of behavior for compounds of uranium (and perhaps other light actinides), but more systematic evidence is necessary. In none of the high-pressure works cited above (except Ref 33) the pressure range used did not allow to observe any saturation of ordering temperatures. In this respect, the present work on UGa$_2$ is pioneering and it will be important to compare with other U compounds when relevant data will be available. Even more importantly, the knowledge of pressure variations of the size of moments is essential, as the discussion above assumes the collapse of U moments causing the $T_C$ downturn in the high pressure range. Precise magnetization measurements on single crystals in the range above 10 GPa would be very difficult. Instead, we propose to study pressure variations of U moments by means of XMCD, which is becoming feasible.

## V. CONCLUSIONS

Studies of the high-quality single crystal of UGa$_2$ have shown that the ferromagnetic ordering temperature of this compound increases with the increasing pressure up to $p$ = 14.2 GPa reaching $T_C$ = 154 K, after which it decreases rapidly till at least $p$ = 15.2 GPa where $T_C$ = 147 K. The increase of pressure up to approximately 20 GPa completely suppresses the magnetic ordering. At such pressures another structure type is, however, adopted. Such $T_C(p)$ dependence indicates that the two-band model is applicable for the description of the character of the 5$f$ states in UGa$_2$.

The low-temperature resistivity of UGa$_2$ is dominated by the electron-magnon scattering with the excitation energy exceeding 60 K, which is comparable to the magnetocrystalline anisotropy in the $ab$-plane. The magnon gap increases with the increasing pressure reaching the value of $\Delta$ = 88 K at $p$ = 8 GPa. The resistivity value of ≈300 µΩcm for the $i//c$ geometry in the paramagnetic state can be attributed to a strong spin-disorder scattering, which also leads to a weak negative slope d$\rho$/d$T$ < 0. The high resistivity and also its negative slope are removed by pressure, assumed to reduce U magnetic moments. This corroborates the fact that the negative slope is due to strong disorder (originating in spin disorder in this case) and not due to e.g. Kondo effect.

The pressure dependences of the magnon gap, of the residual and room-temperature resistivity suggest that the pronounced washout of the magnetic moment of uranium in $UGa_2$ starts at about 8 GPa. The magnetic exchange, which depends on both the magnetic moment of uranium and the hybridization strength, reaches its maximum around 13 GPa. Within the two-band model scenario and the hybridization-induced exchange, the increase of hybridization above 13 GPa cannot compensate for the decrease of the magnetic moment anymore and magnetic exchange strength reflected in $T_C(p)$ starts to decrease.

**Acknowledgements**

This work was supported by the Grant Agency of the Czech Republic under the grant No. P204/10/0330 and P204/12/0150. The work at ITU Karlsruhe has been supported by the EU funding scheme ACTINET-I3. Experiments at the Charles University Experiments were performed at MLTL (http://mltl.eu/), which is supported within the program of Czech Research Infrastructures (project no. LM2011025).